\documentclass[aps,prb,twocolumn,superscriptaddress,10pt]{revtex4-2}
\usepackage[utf8]{inputenc}
\usepackage{libertine}
\usepackage[libertine]{newtxmath}
\usepackage{mathtools,graphicx,microtype,bm}
\usepackage[cal=boondoxo,frak=boondox]{mathalfa}
\usepackage[colorlinks=true,allcolors=blue]{hyperref} 

\begin{document}
\title{Negative dynamic conductance of a quantum wire with unscreened Coulomb interaction}

\author{Bagun S.\ Shchamkhalova}
\email[E-mail:]{s.bagun@gmail.com}
\author{Vladimir A.\ Sablikov}
\email[E-mail:]{sablikov@gmail.com} 
\affiliation{Kotelnikov Institute of Radio Engineering and Electronics, Fryazino Branch, Russian Academy of Sciences, Fryazino, Moscow District, 141190, Russia}

\begin{abstract}
Dynamic conductance and time-of-flight current instability in a quantum wire connected to electron reservoirs under DC bias voltage are studied in the absence of a gate screening the Coulomb interaction of electrons. Due to a strong electron-electron interaction, dramatic rearrangements of the charge density distribution and the potential landscape in the wire occur at a sufficiently high DC bias voltage. The applied voltage is screened mainly near the cathode contact, and an almost flat potential profile is established in the most of the wire. Thus, the size of the region of a population inversion of electronic states greatly increases, and the band of wave vectors that form unstable modes of electronic waves significantly reduces. As a result, the conditions for the occurrence of the time-of-flight instability are greatly facilitated and the negative dynamic conductivity increases. 
\end{abstract}

\maketitle

\section{Introduction}
The idea of a time-of-flight instability dates back to the thirties of the last century in the theory of space charge-limited currents in vacuum it was shown that at a sufficiently large applied voltage, the real part of the admittance becomes negative in certain frequency intervals determined by the injected electrons time of flight in the gap between electrodes~\cite{doi:10.1080/14786443109461701}. However, the main impetus for the development of this idea was given by W. Shockley~\cite{https://doi.org/10.1002/j.1538-7305.1954.tb03742.x}, who showed that the time-of-flight mechanism of negative dynamic conductance (NDC) can be implemented in two-terminal semiconductor structures and created the foundations of the theory of this effect. The development of this direction has led to the creation of a wide family of devices based on the time-of-flight effect, such as IMPATT, BARITT, etc.\ diodes~\cite{sze2006physics}. 

A new surge of interest in time-of-flight effects was provoked by studies of mesoscopic and low-dimensional structures with ballistic transport, since in many situations it is these effects that determine their high-frequency properties. In  low-dimensional structures the role of the electron-electron (e-e) interaction, as is known, greatly increases. As a result of the interaction, not only the spectrum and electronic structure of collective excitations change significantly, but also the spatial distribution of the electron density and electric field in finite systems are strongly rearranged, especially under conditions of strong non-equilibrium that occurs at high applied voltage or current. The e-e interaction leads to the strongest effects in a one-dimensional (1D) or quasi-1D quantum wire (QWr) when metallic gates are located far from the transport channel and therefore weakly screen the Coulomb interaction. Thanks to advances in technology, such structures are now being fabricated and there is considerable interest in their research especially in the case of quantum ballistic or quasi-ballistic regime of the electron transport~\cite{Pogosov_2022}. In this work we study the time-of-flight mechanism of NDC in QWrs of this kind.

The time-of-flight effect and NDC under conditions of a strong inhomogeneity of the electric field, which is formed due to the e-e interaction, was studied for classical three-dimensional semiconductor structures in the ballistic regime of space charge limited current~\cite{Gribnikov}. It has been found that the space charge affects the width of the frequency window in which the dynamic conductance is negative, and shown that the NDC depends significantly on the conditions at the contacts which essentially determines the inhomogeneity of the potential landscape.

In the case of two-dimensional systems, the time-of-flight instability, as far as we know, has been studied only on gated structures in which the potential landscape is controlled by the gate everywhere except for the contact regions. The e-e interaction manifests itself in the formation of plasma waves, the propagation of which determines the dynamic conductivity and leads to specific plasma instabilities.

The most interesting one among them is an instability of a steady-state electron flow in the transistor channel due to the plasma waves reflection from the drain edge of the gated channel, the so-called Dyakonov–Shur instability~\cite{PhysRevLett.71.2465}. However, in structures of this type the time-of-flight instability can also arise. This happens when there is certain part of the structure in which the electrons move ballistically. So, in GaAs HEMT structures, this is a portion of the two-dimensional channel through which the electrons can pass and change the dynamic conductance of the channel~\cite{Ryzhii_2002, Ryzhii_2005,Ryzhii1}. This can also be a region between the gates in a two-dimensional structure with a periodic system of interdigitated gates~\cite{Ryzhii_2008}.

\begin{figure}
   \centerline{\includegraphics[width=0.95\linewidth]{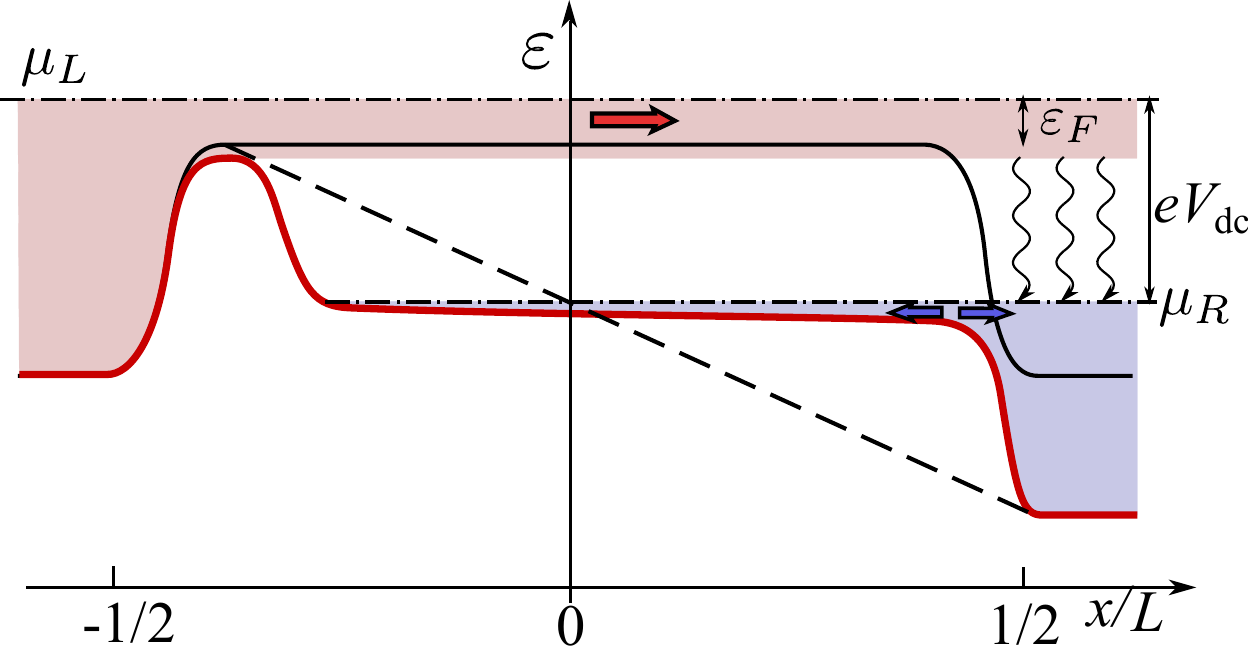}}
    \caption{Schematic energy diagram of a QWr connected to electron reservoirs to which a bias voltage $V_{dc}$ is applied. The thin black line represents the energy diagram of the unbiased QWr. The dotted line shows the potential landscape without the charge accumulated in the QWr. $\mu_{L, R}$ is the chemical potential in the left and right reservoirs, $\varepsilon_F$ is the Fermi energy in the unbiased wire. Wavy lines indicate energy relaxation processes in the reservoir.}
    \label{fig1}
\end{figure}

The time-of-flight mechanism of NDC in 1D QWrs connected to massive electron reservoirs was studied only in the absence of an e-e interaction, when the electric field in the gap between the contacts was assumed to be the same as the field created by the voltage applied to the electrodes and the charges in the QWr were ignored~\cite{Fedichkin,Sablikov1}. This assumption is too rough for 1D systems, especially when the applied voltage is comparable to the kinetic energy, as happens when NDC appears in the ballistic regime. It is known that when an applied voltage is comparable to or exceeds the Fermi energy in an ungated QWr, a strong rearrangement of the spatial distribution of the charge and electric field occurs, which results in the formation of an almost flat potential landscape in most of the wire~\cite{Picciotto,Bachtold}. Our study~\cite{SABLIKOV2003189,SHCHAMKHALOVA200551} showed that this is a result of the development of a specific instability of the charge density distribution, which leads to a screening of the applied voltage near the cathode by a positive charge accumulated in the QWr. As a result, a potential barrier for electrons is formed near the cathode, and an almost flat potential landscape is established in the rest of the wire, as shown in Fig.~\ref{fig1}. Theory of ballistic currents limited by space charge in nanostructures of different dimensionalities,~\cite{beznogov2013theory} showed that this form of potential landscape is a specific and rather general feature of 1D systems. Obviously, the formation of such a landscape leads to a significant increase in the size of the region with an inverse population of electronic states, which is necessary for self-excitation of electron density oscillations. In addition, the set of wave vectors of electrons in the wire is strongly narrowed and, therefore, the amplitude of modes close to resonance increases. Therefore, one can expect a significant easing of the conditions for the emergence of the time-of-flight instability and a notable change in the magnitude of the NDC\@.

In this paper, we find out how the rearrangement of the potential landscape of a QWr due to the strong interaction of electrons affects the conditions for the excitation of the time-of-flight instability and the magnitude of NDC\@. The theory presented here is based on a self-consistent calculation of the potential in the QWr and its contacts to electron reservoirs. We show that the formation of a potential with a relatively narrow barrier and an extended region with an almost flat potential landscape leads to a significant increase in the NDC value and a small decrease in the threshold voltage. At high voltages, the NDC magnitude decreases, and with an increase in the length of the wire, the NDC increases if the length is not too large.

\section{The model of an ungated QWr with leads}\label{S:model}

The QWr connected to the leads is considered as a 1D conductor of a diameter d and a length of 2a, smoothly expanding at the ends, thus providing an adiabatic transition of electrons to the electron 2D or 3D highly conductive reservoirs located at a distance $L$ from each other. The QWr diameter as a function of the coordinate along the QWr is approximated as $d(x)=d[1+\Theta(|x|-a)(|x|-a)^2/R^2]$, with $R$ being the characteristic size of the near-contact region and $x = 0$ in the middle of the QWr. In this paper, we study the linear response of this structure to an AC voltage $V_{ac}\cos(\omega t)$ applied between the reservoirs in the presence of a DC bias $V_{dc}$, due to which the charge and potential distribution in the QWr is strongly rearranged. 

The key role is played by the e-e interaction potential which is known to be the strongest in the QWr. Therefore we neglect the interaction in the reservoirs ($|x|>L/2$). The e-e interaction in the QWr is inhomogeneous due to screening by massive reservoirs. The interaction potential is determined not only by direct Coulomb interaction of electrons but also by the interaction mediated by image charges induced by the interacting electrons on the reservoirs~\cite{PhysRevB.58.13847}. This potential was used to calculate the dynamic conductance of an unbiased QWr~\cite{PhysRevB.58.13847,sablikov2000dynamic} of finite length and non-linear DC conductivity in the presence of a finite bias voltage, including the regime of space-charge limited current~\cite{SABLIKOV2003189,SHCHAMKHALOVA200551}. 

The approach used here to study the dynamic response in the presence of a large enough bias voltage is based on the self-consistent field approximation and the reduction of the Schrödinger equation to an effective one-dimensional equation by integrating over transverse coordinates.
We suppose that the QWr is sufficiently long and narrow so that $L k_F\gg 1$ and $d\,k_F\ll 1$, with $k_F$ being the Fermi wave vector in the middle of the QWr. Thus only the lower subband of the transverse quantization can be considered in the middle part of the channel, though in the near-contact regions several subbands are taken into account. 

Following to the procedure described in Ref.~\cite{SHCHAMKHALOVA200551}, we reduce the problem to an effective 1D Schrödinger equation for the wave functions $\Psi(x,t)$:
\begin{equation} \label{sch}
    i\hbar\frac{\partial \Psi}{\partial t} +\frac{\hbar^2}{2m}\frac{\partial^2 \Psi}{\partial^2 x}- U(x)\Psi-H_{ac}(x,t)\Psi = 0.
\end{equation}
Here $U(x)$ is the effective potential, composed of a transverse quantization energy in the channel and the electrostatic potential of all charges in the system averaged over the transverse wave functions. These are the external charges induced on conducting electrodes by the applied voltage, a positive background charge density and the charge of the electrons distributed in the channel, which is calculated self-consistently with the potential. 

$H_{ac}$ is the Hamiltonian of the electron interaction with the electric field $E(x,t)$ that appears due to the applied AC voltage. We represent it in terms of the vector potential:
\begin{equation} \label{Ha}
    H_{ac}(x,t) = \frac{e\hbar V_{ac}}{2m(\omega+i\eta)}\left[F(x)\frac{\partial }{\partial x} +\frac{1}{2}\frac{\partial F(x) }{\partial x}\right]e^{-i\omega t}+ c.c.,
\end{equation}
where $\eta\to +0$ is standard regularizing factor. 
In what follows we suppose for simplicity that $F(x)$ is determined mainly by the electric field $F(x) = E(x)/V_{ac}$ created by external charges induced by an AC voltage on the electrodes, and ignore the AC electric field due to polarization of an electron system in the QWr. This assumption is justified by results of our study of the electron charge density appearing in the QWr due to an AC voltage in the absence of the DC bias. The calculations in the frame of the Luttinger liquid approach~\cite{PhysRevB.58.13847,sablikov2000dynamic} clearly show that at the frequency comparable or higher than the inverse transit time of bosonic excitations in the QWr, the screening of the external AC electric field is negligible~\footnote{See Fig.~4 in Ref.~\cite{PhysRevB.58.13847} where the AC charge density distribution in the QWr is shown for a set of the frequencies. The charge density significantly decreases with the frequency and almost disappears when $\omega L/v \sim 1$.}. Of course, by writing the Hamiltonian $H_{ac}$ in the form of Eq.~(\ref{Ha}) we somewhat underestimate the renormalization of the charge wave velocity $v$ due to the e-e interaction. However, as can be seen from Fig.~5 in Ref.~\cite{PhysRevB.58.13847} this effect does not change qualitatively the frequency dependence of the dynamic conductance in the range of the order of and below the frequency of the first resonance of plasmons.

Equation~(\ref{sch}) is solved perturbatively. First, we find the wave functions and the potential landscape that are formed under the action of a finite DC bias voltage. The AC response is then calculated by treating $H_{ac}$ as a perturbation. As will be seen, due to the strong rearrangement of the potential landscape with the formation of a barrier near the cathode and almost flat in the rest of the wire, the NDC value and the conditions for its appearance change significantly.

\section{Potential landscape of the biased QWr}\label{S:landscape}
The DC problem is solved in the same way as in Ref.~\cite{SHCHAMKHALOVA200551} so we drop details. The main idea follows from the expression for the effective potential in the 1D equation:
\begin{equation}\label{pot}
    U(x) = \xi(x) + \int dx'G(x,x')[\rho[U(x')]-\rho_b(x')] + \varphi(x) \,,
\end{equation}
where $\xi(x)$ is the transverse quantization energy, $\varphi(x)$ is the potential of external charges induced on the conducting electrodes, $G(x; x')$ is an effective 1D Green function which determines the potential created by the positive background charge and the charge of electrons. The specific form of $G(x; x')$ is found as a result of the integration over the transverse coordinates and in general case is determined by the geometry of the reservoirs and nearby gates. The positive background charge density $\rho_b(x)$ is geometry dependent, for $|x|< a$ it is a constant. The electron charge density $\rho[U(x)]$ is a functional of the 1D potential $U(x)$. We define $\rho[U(x)]$ in the frame of the quasi-classic approximation assuming that $U(x)$ is a sufficiently smooth function. 

Within this approach the electron density is evidently formed on the basis of right- and left-moving states,
\begin{equation}\label{Psi_0}
  \Psi_0^{r,l}(x,E) = \sqrt{\frac{k_0(E)}{k(x,E)}}\exp\left[\pm i\int_{-L/2}^{x}dx'k(x')\right] \,,
\end{equation}
where $k_0(E)=\sqrt{2mE}/\hbar$, $k(x,E)=\sqrt{2m(E-U(x))}/\hbar$, and $E$ is the energy of the corresponding state.

\begin{figure}
\includegraphics[width=0.95\linewidth]{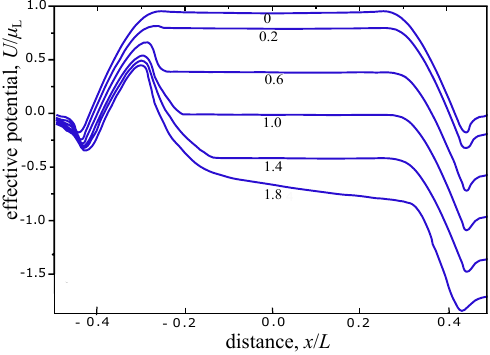}
\caption{Effective 1D potential in a QWr coupled to reservoirs for various bias voltages. The parameters used in the calculations are: $L$=5\,$\mu$m, $d$=20\,nm, $a$=1.25\,$\mu$m, $R$=0.75\,$\mu$m, $Lk_F$= 100, $g\approx 30$. The applied voltage normalized to the Fermi energy in the reservoirs, $V = V_{dc}/\mu_L$,  ($\mu_L$ = 5.5\,m$e$V) is shown below each line.}
\label{profile}
\end{figure}

The self-consistent charge and potential distribution in the QWr under the applied DC voltage are determined from Eq.~(\ref{pot}) numerically for electrodes in the form of two equipotential planes perpendicular to the QWr. In the transition regions between reservoirs and one-mode part of the QWr we take into account the electron charge of several modes of the electron transverse motion. The self-consistent potential $U(x)$ in the QWr and contact regions is shown in Fig.~\ref{profile} for a specific set of parameters and a variety of DC bias voltages. 

The potential landscape $U(x)$ depends strongly on the e-e interaction parameter which can be estimated as $g\approx 2r_s\ln(L/d)$, where $r_s$ is the standard parameter for an electron gas in an infinite 1D QWr. If $g \gg 1$ the main features of the potential landscape can be described analytically~\cite{SHCHAMKHALOVA200551}. The general features of the evolution of the potential landscape with the bias voltage are as follows. 

At low voltages, the charge density in the QWr remains nearly constant and equal to the equilibrium one. This is a result of the large potential energy associated with a charge imbalance due to a high interaction parameter $g$. At voltages smaller than the Fermi energy in the QWr the potential landscape in the QWr remains flat (see line labeled 0.2 in Fig.~\ref{profile}) shifting  down in half of the applied voltage. There are two potential steps at the contact regions where an applied voltage drops equally. 

As the bias voltage exceeds a critical value $V_1$, a potential barrier is formed near the cathode contact while the potential landscape in the QWr remains nearly flat. The critical voltage is estimated as $eV_1/ \mu_L \approx  d^2 k_F^2/4$. In this regime, only the electrons injected from the cathode pass through the QWr, while the electrons coming from the anode are blocked by the barrier (see lines labeled 0.6 and 1.0 in Fig.~\ref{profile}). Nevertheless, the electrons coming from the anode play an important role, since they participate in the maintenance of the nearly flat form of the potential landscape. It is interesting that the main part of the applied voltage drops across the barrier. Under these conditions, the electron energy distribution function is strongly inverted in the large part of the gap between the reservoirs. The electrons injected by the cathode occupy the energy interval $U_m< E < \mu_L$, where $U_m$ is the top of the barrier. The electrons coming from the anode occupy a relatively narrow interval of energy between the bottom of the conduction band in the QWr and the chemical potential of the anode reservoir. 

Another characteristic voltage is $V_2 \approx g V_1/3$. At $V_{dc}>V_2$, the electrochemical potential level in the anode reservoir lies too low and the anode cannot supply enough electrons to screen the electric field in the QWr (line marked 1.8 in Fig.~\ref{profile}). As a result, the electric field in the central part of the QWr increases considerably.

Thus, the main conclusion is that the potential profile of the QWr is strongly rearranged under the large dc bias voltage. The significant part of the applied voltage drops near the cathode contact, while the potential profile over most of the QWr length remains almost flat. This finding is consistent with other results on narrow wires or atomic strings in the presence of high source-drain voltages obtained using various approaches and calculation methods, for a review see Ref.~\cite{lassl2007spin}.

\section{Admittance and NDC}\label{s:Admittance}

The AC current and the admittance are studied using the approach described in Ref.~\cite{Sablikov1}. The AC electron current in the QWr is calculated considering the electron interaction with the AC field as a perturbation. In the first order of the perturbation theory, the electron wave function reads: 
\begin{equation}\label{Psi}
\Psi^{\lambda}(x,t)=\Psi_0^{\lambda}(x,t)+\left[\Psi_+^{\lambda}(x)e^{-i\omega t}+\Psi_-^{\lambda}(x)e^{i\omega t}\right]e^{-iEt/\hbar},
\end{equation}
where $\Psi^{\lambda}$ is the unperturbed wave function of electrons coming from the cathode ($\lambda=R$) or anode ($\lambda=L$); $\pm$ marks components with an energy $E\pm \hbar \omega$.

Of greatest interest is the situation $V_{dc}>V_1$ when  and the electrons coming from the anode are blocked by the barrier in the considered frequency range. In the quasiclassic approximation, the wave function of electrons injected from the cathode $\Psi^R(x,t)$ is formed by the right-moving components $\Psi_0^r$ defined by Eq.~(\ref{Psi_0}). The AC electron current with the frequency $\omega$ reads: 
\begin{equation}\label{jet}
\begin{split}
j^R(x,t)=\frac{i e\hbar}{2m}\left(\Psi_0^{r*}\frac{d\Psi_+^R}{dx}-\Psi_+^R\frac{d\Psi_0^{r*}}{dx}-\Psi_0^r\frac{d\Psi_-^{R*}}{dx} +\Psi_-^{R*}\frac{d\Psi_0^r}{dx}\right. \\- \left. \frac{eV_{ac}}{\hbar\omega}\Psi_0^{r*} \Psi_0^r F(x)\right)e^{-i\omega t}+\mathrm{c.c.}
\end{split}
\end{equation}

The wave function of electrons coming from the anode $\Psi^L(x,t)$ is formed by both components $\Psi_0^r$ and $\Psi_0^l$. These electrons also contribute to the AC current in external circuit though they are blocked by the barrier. The effect is due to AC polarization of these electrons. However, in the considered system this component of the current turns out to be negligibly small since the density of electrons coming from the anode is small in the voltage range of interest for us.

The electric current in the external circuit is found in accordance with the Shockley theorem and has the form:

\begin{equation}
\label{shok}
j(t) = \frac{1}{L}\!\int \limits_{-L/2}^{L/2}\!dx F(x)\int\limits_{U_m}^{\infty}\! dE g(E)\,f(E) j^R(x,t)+C \frac{dV_{ac}(t)}{dt}\,,
\end{equation}
where $C$ is interelectrode capacitance, $g(E)$ is the density of states and $f(E)$ is the Fermi distribution functions in the cathode reservoir, 
$f(E)=[1+e^{(E-\mu_L)/kT}]^{-1}.$ 
The current calculated in this way is used to determine the admittance. For simplicity, further calculations are carried out for $T\to 0$.

The final expression for the admittance $Y(\omega)$ normalized to the conductance quantum $e^2/h$ has the following form:

\begin{equation}
  Y(\omega)=Y_+(\omega) + Y_-(\omega) - i \frac{h \omega C}{e^2}\,, 
\end{equation}
where $Y_{\pm}$ corresponds to the higher and lower energy sideband. They are given by
\begin{widetext}
\begin{equation}\label{adm_p}
\begin{split}
Y_+(\omega) =&\frac{1}{\hbar\omega}\int\limits_{-L/2}^{L/2}dxF(x)\int\limits_{U_m}^{\infty}dE \left\{[f(E) -f(E+\hbar\omega)]
\frac{k_++k}{\sqrt{kk_+}}e^{i(S_+-S)}\int\limits_{-L/2}^{x}dx'F(x')e^{-i(S_+-S)}\right. \\
&+\frac{k_+-k}{\sqrt{kk_+}}\left[f(E)e^{-i(S_++S)}\int\limits_{x}^{L/2}dx'F(x')\frac{k_+-k}{\sqrt{kk_+}}e^{i(S_++S)}
-\left. f(E+\hbar\omega)e^{i(S_++S)}\int\limits^{L/2}_{x}dx'F(x')\frac{k_+-k}{\sqrt{kk_+}}e^{-i(S_++S)}\right]\right\}\,,
\end{split}
\end{equation}
\begin{equation}\label{adm_m}
\begin{split}
Y_-(\omega) = &-\frac{1}{\hbar\omega}\int\limits_{U_m}^{U_m+\hbar\omega}dE f(E) \left\{\int\limits_{x_-}^{L/2}dx F(x)\left[\frac{k+k_-}{\sqrt{kk_-}}e^{i(S-\tilde S_-)}
\int\limits_{x_-}^{L/2}dx'F(x')\left\{\frac{k_-+k}{\sqrt{kk_-}}e^{-i(S-\tilde S_-)}+\frac{k-k_-}{\sqrt{kk_-}}e^{-i(S+\tilde S_-)}\right\}\right.\right. \\
&+\left.\left.\frac{k-k_-}{\sqrt{kk_-}}e^{-i(S_-+S)}\int\limits_{x_-}^{x}dx'F(x')\frac{k-k_-}{\sqrt{kk_-}}e^{-i(S_-+S)}
- 4F(x)\left(\frac{\cos{\tilde S_-}}{k_-}+ \frac{\sin{\tilde S_-}}{k_-}\right)\right]\right\}\,,
\end{split}
\end{equation}
\end{widetext}
where  $k_\pm=\sqrt{2m(E-U(x)\pm \hbar\omega)}$, $S(x)=\int_{-L/2}^x dx'k(x')$, $S_+(x)=\int_{-L/2}^x dx'k_+(x')$, $\tilde{S}_-(x)=\int_{x_-}^x dx'k_-(x') -\pi/4$.
The lower sideband has a classical turning point $x=x_-$ for $E <U_m +\hbar\omega$, to the left of which the quantity $k_-(x)$ becomes imaginary; $x_-$ is a root of the equation $$E +\hbar\omega - U(x_-)=0.$$

The electron flow from the cathode to the anode with emission and absorption of photons gives the main contribution to the AC current. The maximum of the potential barrier $U_m$ is in the single-mode part of the wire at some distance from the cathode reservoir. The electrons moving from the cathode with an energy $E< U_m$ are returned. Their contribution to the admittance is small when the virtual cathode $U_m$ is located near the cathode reservoir. 

\begin{figure}
\includegraphics[width=0.95\linewidth]{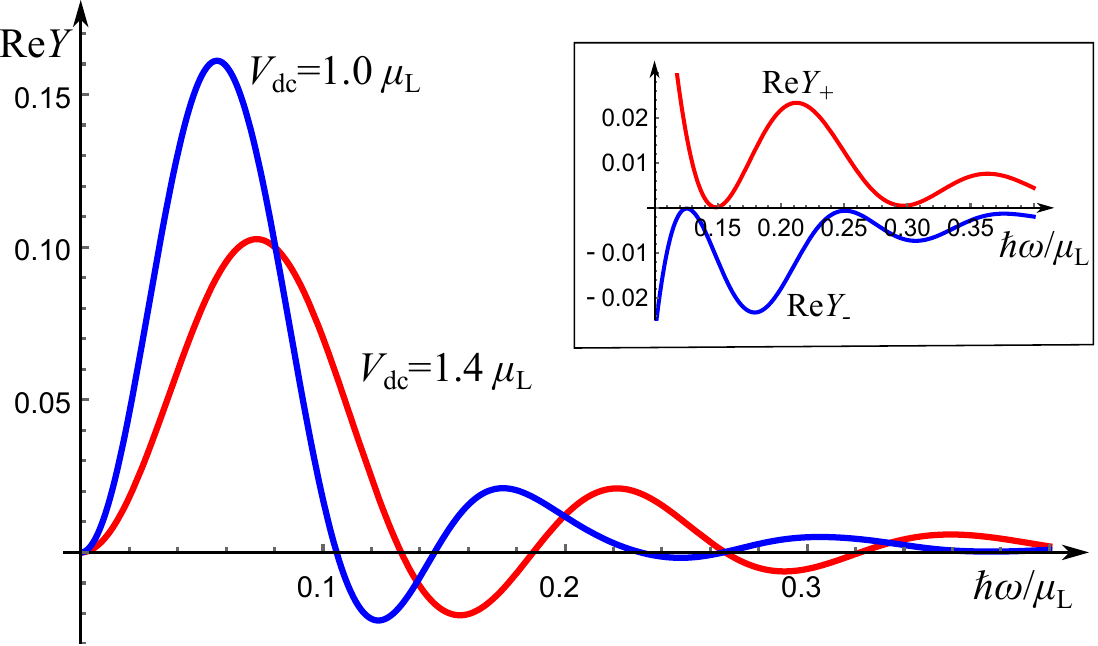}
\caption{Real part of the admittance as a function of frequency calculated for parameters of Fig.~\ref{profile} for two voltages $V_{dc}=1.0\,\mu_L$ and $V_{dc}=1.4\,\mu_L$. Inset shows the frequency dependency of the real parts of $Y_+$ and $Y_-$ for $V_{dc}=1.4\,\mu_L$.}
\label{adm_real}
\end{figure}

Of most interest is the real part of the admittance and its dependence on the frequency, the bias voltage, and the QWr length. As a function of the frequency, $\mathrm{Re}Y(\omega)$ exhibits an oscillatory behavior with amplitude decreasing with the frequency and changing the sign, as expected in the theory of the time-of-flight instability. Figure~\ref{adm_real} shows an example of such a dependence calculated for the same parameters as in Figure~\ref{profile} and two values of the bias voltage in the regime, when  $V_1<V_{dc}<V_2$ and the potential landscape is practically flat over most of the QWr. The dynamic conductance becomes negative in frequency ranges associated with the transit time of electrons.

It is interesting to consider the contributions of both components $Y_+$ and $Y_-$ corresponding to the upper and lower sidebands into the admittance. They describe respectively processes with absorption and emission of quanta $\hbar \omega$. Fig.~\ref{adm_real} clearly shows that both components are oscillating functions of $\omega$ with different periods. This can be interpreted as a result of the difference between the electron velocities in the upper and lower side bands. Thus, along with the presence of population inversion, the realization of NDC requires a difference in effective velocities of the electron density waves in the upper and lower side bands.

In the large voltage regime $V_{dc}>>V_2$, the applied voltage drops mainly in the QWr and the electric field in the QWr becomes large (see curve 1.8 in Fig.~\ref{profile}). We considered this case separately, approximating the QWr potential profile with a linear function, and found that the value of the real part oscillates with frequency similar to that in Fig.~\ref{adm_real}, with decreasing amplitude.

The calculations we carried out make it possible to find out how the rearrangement of the potential landscape with the formation of a barrier and an almost flat section in the QWr manifests itself in a high-frequency response and the formation of NDC. To this end, we compared our results for a realistic model, which takes into account the rearrangement of the potential landscape, with the results of calculations for an idealized model, in which the potential landscape is approximated by a linear function. We omit the details of this calculation. 

The most interesting and practically important quantity characterizing the time-of-flight effect is the maximum value of the NDC and the frequency at which this maximum is reached. The results of the study of these quantities are presented below. 

\begin{figure}
    \includegraphics[width=0.95\linewidth]{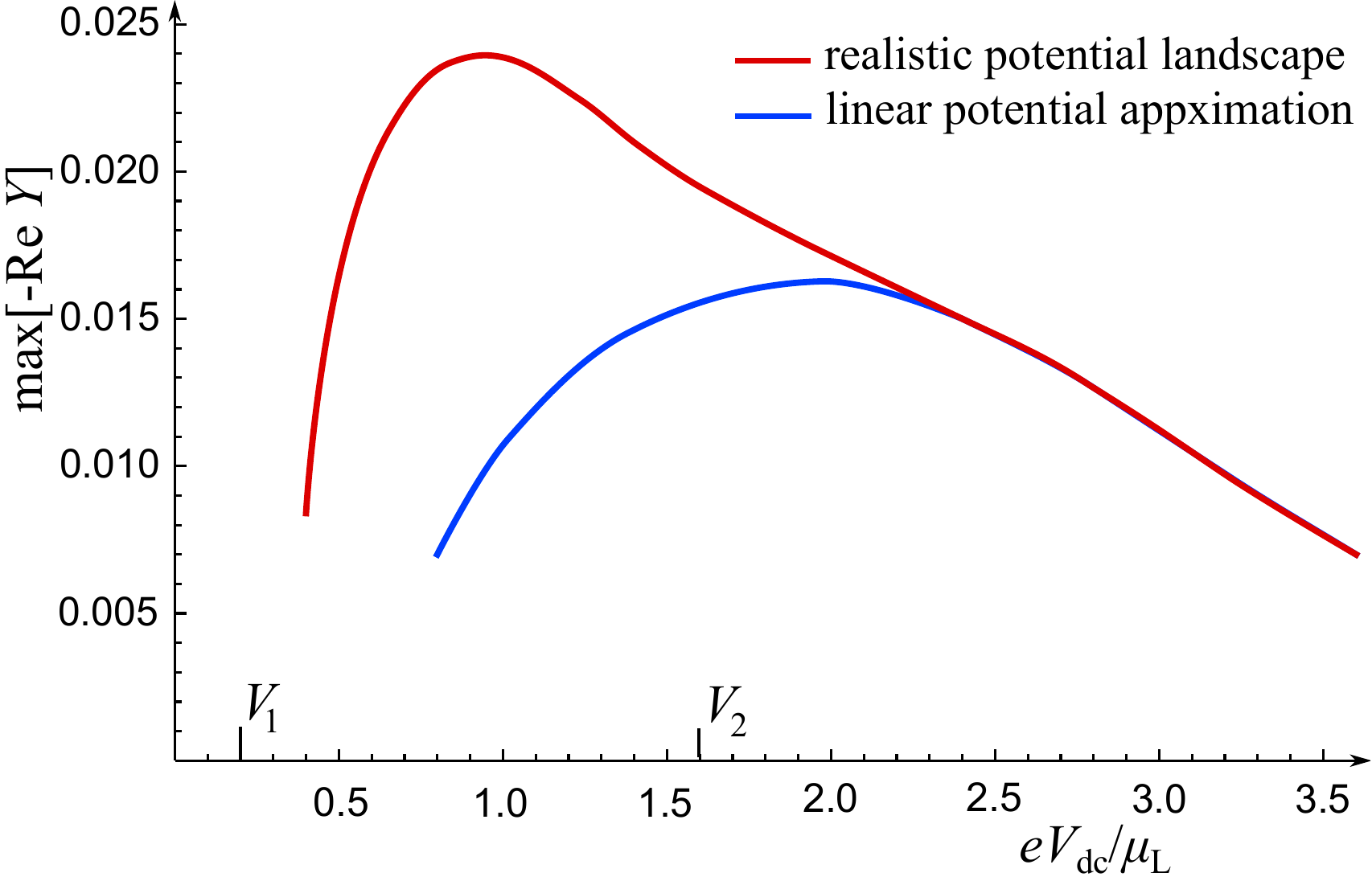}
    \caption{Maximum NDC as a function of the DC bias voltage for the realistic potential landscape of the QWr (red line) and the linear potential approximation. The calculations were carried out for parameters of Fig.~\ref{profile}}
    \label{adm_max}
\end{figure}

Figure~\ref{adm_max} shows the dependence of the maximal value of the NDC, $\mathrm{max}[-\mathrm{Re}Y]$, on the bias voltage for the realistic potential landscape and the linear approximation for the same parameters of the electronic system of the QWr. It is clearly seen that in the situation when the potential landscape of the QWr is rearranged, the NDC appears at a lower bias voltage, than in case of the potential approximated by a linear function. Then, as the voltage increases in the range $V_1<V_{dc}<V_2$ when the restructuring of the potential landscape takes place, the NDC in the realistic model becomes several times larger. 

\begin{figure}
\includegraphics[width=0.9\linewidth]{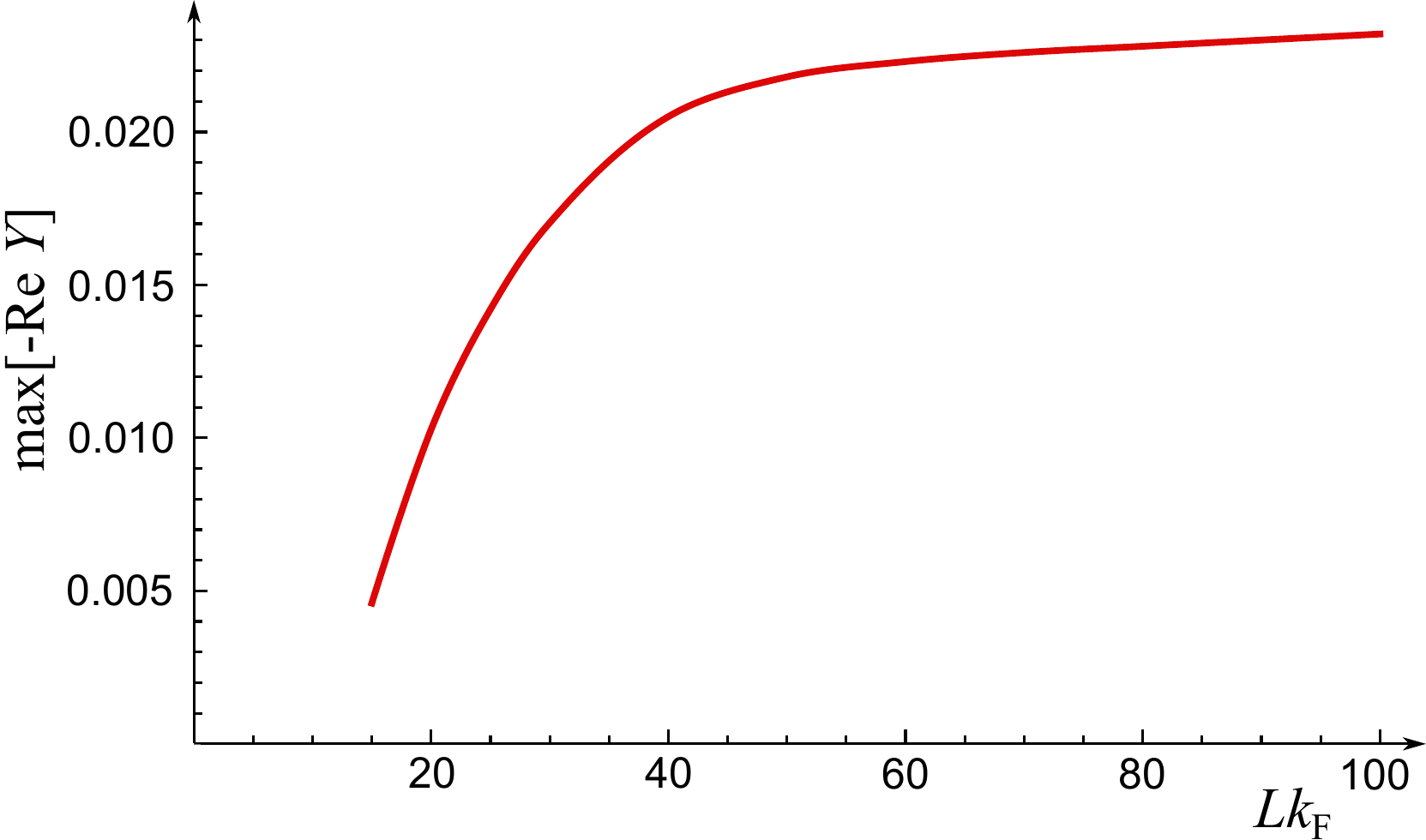}
\caption{Maximum NDC as a function of the QWr length in the regime of the rebuilt potential landscape. The calculations were carried out for parameters of Fig.~\ref{profile} and the bias voltage $V_{dc}/\mu_L =1.4$.}
\label{max_L}
\end{figure}

The dependence of the maximum NDC on the QWr length is shown in Fig.~\ref{max_L} for a given bias voltage in the range where the potential landscape is rearranged. 
The increase in NDC with an increase in the QWr length at a short length seems to be due to an increase in the size of the volume where the electrons interact with the AC electric field. Further saturation of this dependence with increasing $L$ is associated with the flat potential landscape of most of the QWr.

\section{Discussion and concluding remarks}\label{S:conclusion}

We have studied the dynamic conductivity and the time-of-flight instability of the current in an ungated QWr at a sufficiently high bias voltage, when the potential landscape in the wire is rearranged due to the strong e–e interaction and charge redistribution. The resulting potential landscape is characterized by the presence of a barrier near the cathode contact and an extended almost flat area. It is this potential landscape that is realized in real ungated QWrs. Owing to this potential distribution in the wire, the flow of electrons injected from the cathode is formed in the energy band that lies much higher than the energy of the states filled with electrons coming from the anode. The presence of such a flow was found in experiments on the scattering of electrons in normally pinched-off QWr formed in GaAs/AlGaAs heterostructures~\cite{Novoselov}. Thus, a situation arises when the population of electronic states in the QWr is strongly inverted and the injected electron flow is formed by a set of wave functions with a relatively small dispersion of wave vectors. We have found that in this case the conditions for the emergence of the time-of-flight instability are greatly facilitated and the value of the NDC increases.

Within the framework of the quasi-classical approach used in here, the mechanism of a formation of the NDC is as follows. Under the action of an AC electric field, a traveling wave of electron density is formed in the QWr. Due to the inertia of the movement of particles, only a part of the electrons moves against the AC electric field, receiving energy from it. Other electrons move along the field, transferring their energy to it. Accordingly, two components of the electron current are formed, one due to the absorption of a quanta $\hbar \omega$, and the other due to the emission of a quanta $\hbar \omega$. The first makes a positive contribution to the conductivity, and the second makes a negative one, as described by Eqs.~(\ref{jet}) and (\ref{adm_p}),~(\ref{adm_m}). We have found that both components oscillate with frequency, Fig.~\ref{adm_real}, with different periods due to electron dispersion and inhomogeneity of the potential landscape. As a result, in some frequency intervals, the total conductance becomes negative. 

Our calculations have shown that, due to the rearrangement of the potential landscape, the maximum value of NDC increases by several times. At the same time, the frequency of the NDC maximum also increases.

The effects of the potential rearrangement  in an ungated wire found here arise really at a low bias voltage. The potential rearrangement begins at a voltage $V_1\sim\mu_F (d\,k_F)^2/4$, which is estimated at the level $V_1\sim$ 1 -- 2\,m\textit{V} for structures of the GaAs/AlGaAs type. The strongest effects appear at $V\lesssim V_2$, which exceeds $V_1$ by a factor of about $2r_s \ln(L/d)$. Thus, instability can manifest itself at voltages of the order of several m\textit{V}.

The instability of the electronic system that occurs in a QWr at a sufficiently large bias can, of course, lead to generation of a microwave radiation of low power. But there is another aspect of the manifestation of the instability associated with its influence on the nonlinear transport due to the rectification of the alternating current. These effects require a further nonlinear analysis.

The idea of strong rearrangement of the potential landscape of the QWr and possible generation of a high-frequency signal can be useful to understand transport anomalies observed in many studies: see Refs.{\cite{Kumar,condmat7030049} and references therein. New prospects for the manifestation of the processes of the rearrangement of potential relief and their effects on time-dependent processes in in nanostructures open up in connection with growth of interests in space charge-limited currents in such systems in recent years~\cite{10.1063/5.0042355}.

\begin{acknowledgments}
This work was carried out in the framework of the state task for the Kotelnikov Institute of Radio Engineering and Electronics.
\end{acknowledgments}
    
\bibliography{t-f-instability}
\end{document}